\documentclass[twocolumn,conference]{IEEEtran}

\usepackage[style=ieee,backend=bibtex,sortcites=true,sorting=nyt,hyperref=false,backref=false]{biblatex}
\usepackage{microtype}

\usepackage[T1]{fontenc}
\usepackage[utf8]{inputenc}
\usepackage[english]{babel}
\usepackage{graphicx}
\usepackage[font=small,justification=centering,labelsep=period]{caption}
\usepackage{ifthen}
\usepackage{amssymb}
\usepackage{dsfont}
\usepackage{amsmath}
\usepackage{xcolor}
\usepackage{tabularx}
\usepackage{url}
\usepackage{amsmath}
\usepackage{multirow}
\usepackage{flushend}

\usepackage{hyperref}
\usepackage{booktabs}
\usepackage{fancyhdr}
\usepackage{lineno}
\usepackage{setspace}

\usepackage{soul}

\bibliography{bib-sports.bib}

\newboolean{showcomments}
\setboolean{showcomments}{true}
\ifthenelse{\boolean{showcomments}}
{ \newcommand{\mynote}[3]{
		\fbox{\bfseries\sffamily\scriptsize#1}
		{\small$\blacktriangleright$\textsf{\emph{\color{#3}{#2}}}$\blacktriangleleft$}}}
{ \newcommand{\mynote}[3]{}}
\usepackage{etoolbox}
\patchcmd{\pprintMaketitle}
{\ifvoid\absbox\else\unvbox\absbox\par\vskip10pt\fi}
{\ifvoid\absbox\else\clearpage\unvbox\absbox\par\vskip30pt\fi}
{}{}
\patchcmd{\pprintMaketitle}
{\hrule\vskip12pt}
{}
{}{}
\patchcmd{\pprintMaketitle}
{\hrule\vskip12pt}
{}
{}{}
\appto{\pprintMaketitle}{\clearpage}

\graphicspath{{figures/}}
\begin{document}

\title{National Bias of International Gymnastics Judges during the 2013--2016 Olympic Cycle}

\author{
\IEEEauthorblockN{Sandro Heiniger}
\IEEEauthorblockA{Universität St.Gallen \\ Switzerland \\ sandro.heiniger@unisg.ch}
\and
\IEEEauthorblockN{Hugues Mercier}
\IEEEauthorblockA{Universit\'{e} de Neuch\^{a}tel \\ Switzerland \\ hugues.mercier@unine.ch}
}

\maketitle

\begin{abstract}
 
National bias in sports judging is a well-known issue and has been observed in several sports: some judges have the tendency to give higher marks to athletes of the same nationality. In this work, we study the national bias of international gymnastics judges during the 2013--2016 Olympic cycle. As opposed to prior work, our analysis leverages the intrinsic variability of the judging error based on the performance level of the gymnasts for each apparatus and discipline. 
The magnitude of the national bias varies across judges, nations and disciplines. While acrobatics and trampoline do not exhibit any bias, we observe considerable bias by aerobics, artistics and rhythmics judges. In aerobic and artistic gymnastics this bias further increases for the best athletes competing in the finals. On the positive side, we show that judges are unbiased against direct competitors of their own gymnasts. Our approach could easily be applied to other sports that incorporate a judging panel and objective judging guidelines. It could help sports federations and the public at large understand the extent of national bias and identify particularly prone judges or nations.

\end{abstract}


\section{Introduction}\label{sec:Introduction}

Judging sports competitions is a challenging task. Despite a myriad of technological advances, judges must assess the performance of athletes live, surrounded by thousands of cheering spectators, and according to hundreds of instructions specified in scoring regulations. These evaluations anoint international champions and Olympic medalists, and all the involved parties -- athletes, coaches, fans, officials, sponsors -- have a vested interest in having accurate and fair judges.

In this article we focus on the fairness aspect of judging. Fair judges grade every performance as accurately as possible without introducing any subjective biases into their evaluation. There are many well-known biases in sports judging, and in most cases it is impossible to determine whether theses biases are intentional or not. The most studied and discussed bias in sports is \emph{national bias}. National bias comes in two flavors: judges can favor athletes of the same nationality, and at the same time penalize their close competitors. National bias was shown to exist in many sports including figure skating \cite{Campbell:1996, Zitzewitz:2006, Zitzewitz:2014}, ski jumping \cite{Zitzewitz:2006}, rhythmic gymnastics \cite{Popovic:2000}, artistic gymnastics \cite{Ansorge:1988, Leskovsek:2012}, Muay Thai boxing \cite{Myers:2006}, diving \cite{Emerson:2009} and dressage \cite{Sandberg:2018}. 

The analytical approaches to identify national bias are manifold. Some analyses use sign tests \cite{Ansorge:1988, Campbell:1996, Popovic:2000} or permutation tests \cite{Emerson:2011}, whereas other models use linear regressions \cite{Zitzewitz:2006, Myers:2006, Emerson:2009, Leskovsek:2012, Sandberg:2018}. \textcite{Zitzewitz:2006} links the appearance of national bias to the selection procedure of judges and shows that in figure skating, national bias increases with the importance of the event. He also reveals vote trading in figure skating, where judges reinforce national bias of other judges, and compensation effects in ski jumping, where judges weaken national bias of other judges.

\subsection{Estimating national bias by modeling the heteroscedasticity of judging errors in gymnastics }
Judging in gymnastics is a noisy process and does not rely on comprehensive technical assistance. Athletes are evaluated live by panels of judges, and the final scores aggregate the individual marks given by these judges. This aggregation process typically uses the median or the trimmed mean to remove outliers and improve the accuracy of the overall evaluation. Judges within a panel grade the same performance but rarely agree on a single grade. They may have individual preferences and interpret the scoring regulations differently, but more importantly they overlook or misinterpret elements or mistakes of a performance. Even experienced judges with sophisticated cognitive judging strategies and sensorimotor experiences have a low error detection rate \cite{Flessas:2015}. This leads to an inevitable element of subjectivity and randomness in the judging process which introduces an \emph{intrinsic judging error variability} in the marks given by each judge.

This article is the second of a series of three articles on sports judging. In the first article \cite{MH2018:gymnastics}, we showed that the intrinsic judging error variability in international gymnastic competitions depends on the quality of the performance of the gymnasts: judges are more precise judging the best athletes than mediocre ones. We modeled and quantified the standard deviation of this judging error very accurately with heteroscedastic random variables for each apparatus and discipline using data from continental and international gymnastic competitions held during the 2013--2016 Olympic cycle. The approach, which we leverage in this work, is summarized in Appendix~\ref{app:methodology}.  More precisely, we integrate the knowledge of the distribution of the intrinsic judging error variability into our national bias model. Contrary to previous regression-based analyses in gymnastics, e.g. by \textcite{Leskovsek:2012}, this allows us to quantify the national bias not only on a nominal level but also in terms of the intrinsic variability of judging errors. This is essential to evaluate the severity of the bias. After all, a national bias of 0.2 in favor of an athlete deserving 5.7, with 10.0 being the maximum possible mark, is relatively small and has no impact at the top of a competition, whereas the same bias in favor of an athlete deserving 9.7 is appalling and can change the medalists at the Olympic Games. Furthermore for the best athletes, a judge with a large nominal national bias will likely be accused of intentional misjudgement, whereas even a small nominal national bias can remain very large compared to the other sources of judging errors and impact the ranking of the athletes. We believe that judges are, deliberately or unconsciously, aware of this, and that the national bias follows in magnitude this intrinsic variability.

\subsection{Summary of our results}

We analyze national bias of execution and artistry judges during the 2013--2016 Olympic cycle in the five gymnastic disciplines: acrobatics, aerobics, artistics, rhythmics and trampoline. We show that national bias varies by discipline, by nationality, and by judge. While acrobatics and trampoline judges do not exhibit any national bias in favor of their own athletes, we reveal significant bias in aerobic, artistic and rhythmic gymnastics. Judges in men's artistic gymnastics are the worst offenders: their average over-scoring for their compatriots is half the intrinsic judging error variability, and increases to two thirds of the intrinsic variability in favor of their best compatriots competing in the finals. We also show that judges in all disciplines are unbiased against direct competitors of their own athletes.

The bad news is that at the individual level, the national bias of some judges is two and even three times larger than the intrinsic error variability of an average judge. More plainly: the national bias of these judges is two to three times larger than all the sources of errors of an average judge. Even though we can never entirely rule out inadvertence or unlucky coincidence, the magnitude and statistical significance of this bias is too large for the Fédération Internationale de Gymnastique (FIG) to leave it unaddressed: for most of these judges we obtain p-values of less than one percent, which is a strong refutation to possible statistical anomaly. The good news is that in only one finals (an international competition held in 2013) has this national bias modified the podium in artistics gymnastics. The fact that this did not occur more frequently is a testament to the efforts made by the FIG to avoid same-nationality judges in the finals whenever possible, and to the aggregation mechanisms excluding the worst and best marks from the judging panels. In the conclusion, we propose steps to further decrease the impact of national bias, especially in all-around finals in artistic and rhythmic gymnastics where it is difficult to avoid same-nationality evaluations. 

Note that in the third article of the series \cite{HM2018:heteroscedasticity}, 
we show that the intrinsic judging error variability has the same characteristic heteroscedastic shape in other sports using judging panels and marks within a finite range. The integration of this behaviour could provide an improved national bias analysis in all these sports.

\section{Judging in Gymnastics} \label{sec:Judging}

The five main gymnastics disciplines recognized by the Fédération Internationale de Gymnastique (FIG) are artistic gymnastics, rhythmic gymnastics and trampoline, which are Olympic sports, and aerobic gymnastics and acrobatic gymnastics, which have a world championship held every two years. Gymnastics disciplines have different apparatus and competition formats. Acrobatic gymnastics routines are performed in pairs or in groups; men, women and mixed competitions are held. Aerobic gymnastics features individual and group routines; group routines can be mixed or split by gender. Artistic gymnastics is split by gender; men compete on six apparatuses (floor exercise, parallel bars, horizontal bar, pommel horse, still rings and vault) and women compete on four (balance beam, floor exercise, uneven bars and vault). Rhythmic gymnastics is only practiced by women; it includes individual routines with one apparatus (ball, club, hoop or ribbon) and group routines with one or two apparatus. Trampoline is split by gender, but men and women compete in the same events: individual and synchronized trampoline, double mini-trampoline, and tumbling.

Gymnastics competitions typically consist of a qualifying round followed by a final regrouping the best qualifiers. A gymnastic routine at the international level is evaluated by panels of judges focusing on the difficulty, artistry and execution components of the performance. The final scores and the rankings of the gymnasts are obtained by aggregating the marks from the judges. The number of panels, the number of judges per panel and the aggregation procedure vary per discipline.

In this article we focus on execution judges in all the disciplines, with the additional inclusion of artistry judges in acrobatic and aerobic gymnastics\footnote{Besides acrobatic and aerobic gymnastics, the other disciplines do not feature artistry judges.}. After the completion of a routine by the gymnast, each execution judge in the panel grades the performance with a mark from $0$ to $10$ at steps of $0.1$. The evaluation of a gymnastics routine is based on precise guidelines specified in the Code of Points of each discipline and apparatus\footnote{The gymnastics rules, including the Codes of Points, are available at https://www.gymnastics.sport/site/rules/rules.php.}. All the disciplines with the exception of trampoline can include two reference judges who evaluate performances based on the same criteria as the execution and artistry panel judges, but with increased decision weight if they strongly diverge from the panel.

\section{Data} \label{sec:NB_Data}
The data, provided by the FIG\footnote{Consult www.gymnastics.sport.} and Longines\footnote{Consult www.longines.com.}, encompasses 21 international and continental competitions held between 2013 and 2016, including the 2016 Rio Olympic Games. Table~\ref{tab:NB_Data} shows the size of the dataset by discipline after the following preprocessing. The number of marks depends on the number of performances in the dataset and the size of the judging panels, the latter ranging from four to nine judges. We analyze artistic and execution marks, including those from reference judges, but  exclude marks for the difficulty component. We also exclude synchronized trampoline since its panels are not amenable to analysis due to their small size. We restrict the dataset to performances with a median panel mark of at least $7.0$ to exclude noisy aborted or poorly executed routines such as gymnasts stepping outside the trampoline boundaries or falling from the horizontal bar mid-routine. This excludes $9.9$\% of the original data points. 

\begin{table}
	\footnotesize
	\centering 
	\def\arraystretch{1.2}
	\begin{tabularx}{\columnwidth}{Xcccc} 
		\toprule
		&  Nb. of & Nb. of & Nb. of same  & Nb. of direct\\
		Discipline & routines &  marks & nationality marks & competitors\\ 
		\midrule 
		Acrobatics & 714 & 4'874 & 257 (5.3\%) & 843 (17.3\%) \\
		Aerobics & 921 & 6'396 & 200 (3.1\%) & 757 (11.8\%) \\
		Artistics (M)& 7'120 & 46'748 & 909 (1.9\%) & 3'006 (6.4\%) \\
		Artistics (F)& 3'545 & 23'515 & 522 (2.2\%) & 1'694 (7.2\%) \\
		Rhythmics  & 2'636 & 17'673 & 405 (2.3\%) & 1'297 (7.3\%) \\
		Trampoline & 1'483 & 7'278 & 343 (4.7\%) & 833 (11.4\%) \\[0.4ex]
		\bottomrule
	\end{tabularx} 
	\caption{Sample size by discipline.} 
	\label{tab:NB_Data} 
\end{table}

Since we are interested in the raw marks reported by judges, we disregard penalties outside their jurisdiction and post-evaluation aggregation. We do not distinguish between reference and regular panel judges, which are all part of a single and enlarged panel for our analysis. This is further justified from our previous work \cite{MH2018:gymnastics} showing that reference and regular panel judges exhibit a similar intrinsic judging error variability. We also show in \cite{MH2018:gymnastics} that artistry and execution judges in acrobatic and aerobic gymnastics behave similarly.  

Table \ref{tab:NB_Data} also shows the number of marks given to athletes of the same nationality as the judge as well as their direct competitors. The share of same-nationality marks is around \mbox{$2\%-5\%$}, depending on the discipline. Most same-nationality marks occur in all-around finals where they are difficult to avoid. We define direct competitors as gymnasts ranked immediately ahead or behind an athlete of the same nationality as the judge. A gymnast can have different competitors in qualifying rounds and finals, and the number of direct competitors can be higher than two if multiple gymnasts obtain the same score.

\section{Methods}\label{sec:NB_Methods}
We develop our regression model starting with the mathematical essence of judging: the mark $s_{p,j}$ of performance $p$ by judge $j$ is expressed as 
\begin{equation}\label{eq:basic_formula}
	s_{p,j}=\lambda_p +  \epsilon_{p,j}
\end{equation}
where $\lambda_p$ is the unknown true quality of the performance and $\epsilon_{p,j}$ is a random error term to be discussed later. In gymnastics, the true performance level $\lambda_p$ is called the \emph{control score} $c_p$ and typically obtained by technical committees using post-competition video reviews. Although one could argue that the control score is still an approximation, it is assuredly very close to the true performance quality $\lambda_p$ in practice.  Since the control scores are unavailable for our analysis, we assume that the judging panel is large enough to provide a reasonably good approximation of the true performance level. We thus use the median panel mark, redefine the control score as $c_p\triangleq \underset{j}{\text{med}}(s_{p,j})$, and use this as an estimation for $\lambda_p$. We use the median  because it is less prone to be affected by biased and erratic judges compared to other aggregation measures based on means or trimmed means. The implications of using the median as the control score are discussed in Section~\ref{sec:median}.

Refining this simple model, we consider the general tendency $\mu_j$ of a judge who consistently applies the judging regulations too harshly or too generously\footnote{$\mu_j$ is generally very small, i.e., below 0.02, however for some judges this deviation reaches 0.2 points.}. We then include the national bias of a judge in favor of an athlete with the same nationality with the binary indicator $\mathds{1}_{\text{SN}}$. The extent of the bias is determined by the parameter $\beta_{\text{SN}}$ and estimated by the regression model. National bias does also occur by penalizing direct competitors of same-nationality athletes; we integrate this into the model with the binary indicator $\mathds{1}_{\text{COMP}}$ and the parameter $\beta_{\text{COMP}}$. Our improved model becomes
\begin{equation}\label{eq:update_formula}
	s_{p,j}=c_p + \mu_j + \beta_{\text{SN}} \cdot \mathds{1}_{\text{SN}} + \beta_{\text{COMP}} \cdot \mathds{1}_{\text{COMP}} + \epsilon_{p,j}.
      \end{equation}
      
\begin{figure}
\centering
\includegraphics[width=\columnwidth]{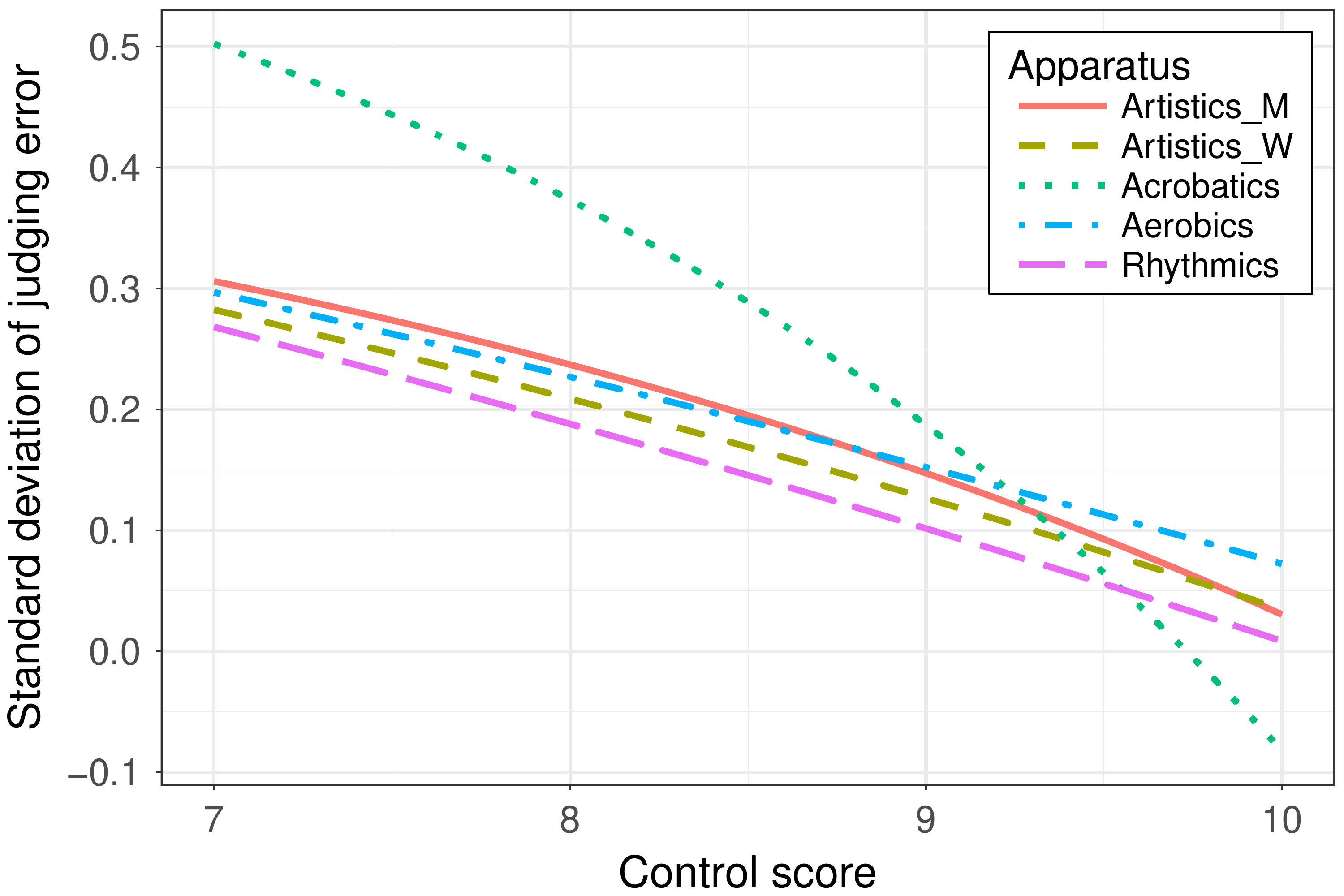}
\caption{Intrinsic discipline judging error variability $\hat{\sigma}_d(c_p)$: fitted standard deviation of judging error as a function of the control score per discipline.}
\label{fig:sigma_DIS}
\end{figure}

\begin{figure}
\centering
\includegraphics[width=\columnwidth]{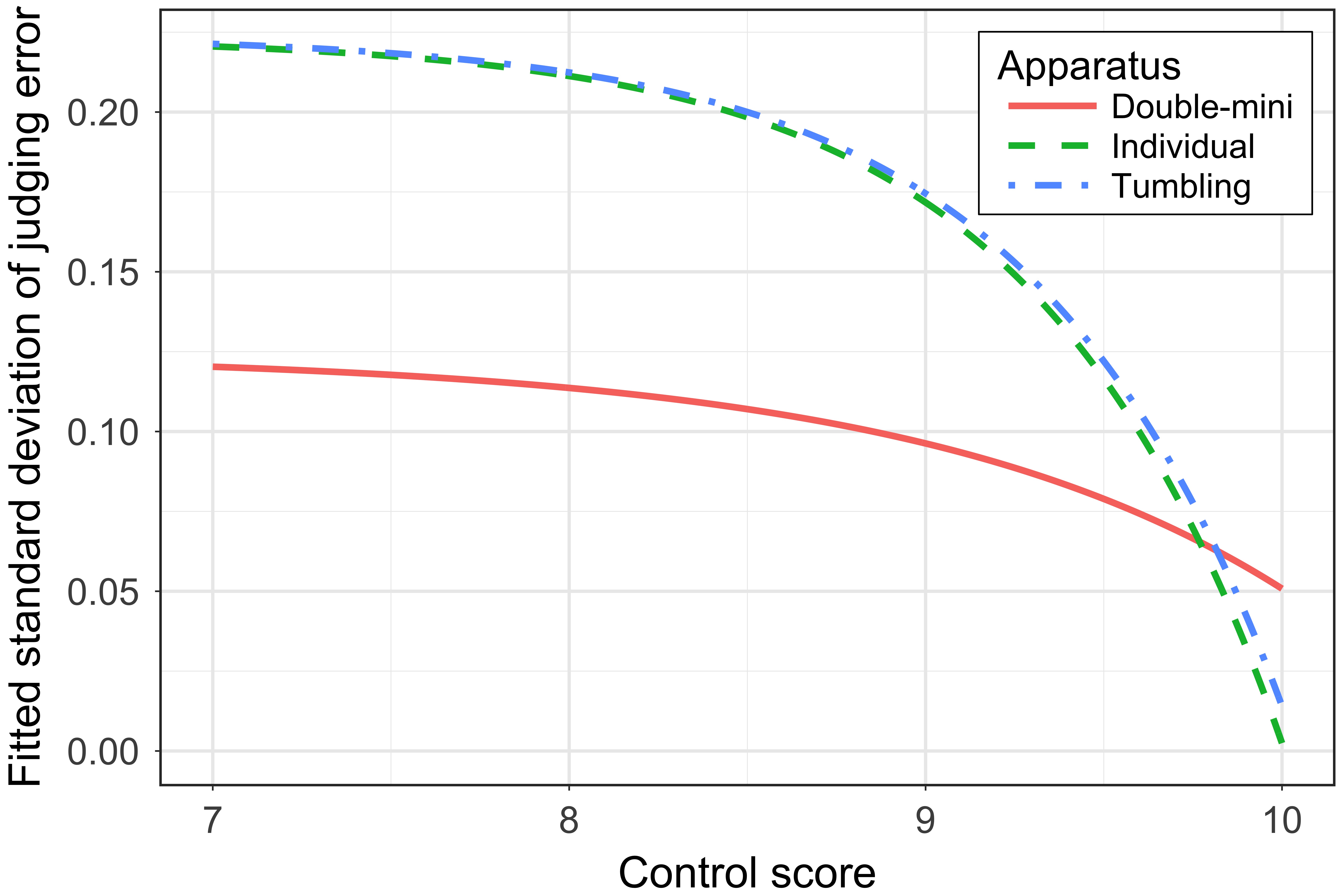}
\caption{Intrinsic discipline judging error variability $\hat{\sigma}_d(c_p)$: fitted standard deviation of judging error as a function of the control score per apparatus in trampoline.}
\label{fig:sigma_APP}
\end{figure}

In our first article \cite{MH2018:gymnastics} we show that the judging error for a control score $c_p$ in discipline $d$ is a random variable with mean 0 and standard deviation $\sigma_d(c_p)$ where $\sigma_d(c_p)$ is an exponential function. This function $\sigma_d(c_p)$, which we call \emph{intrinsic discipline judging error variability}, quantifies the errors made by an average judge. It is specific to each discipline, each with its peculiarities and judging challenges. Appendix~\ref{app:methodology} briefly explains how to estimate $\hat{\sigma}_d(c_p)$ from the data. Figure~\ref{fig:sigma_DIS} shows $\hat{\sigma}_d(c_p)$ as a function of the control score $c_p$ for all the disciplines except trampoline, and Table~\ref{tab:magsigma} shows a few numerical values for men's artistic gymnastics to further illustrate the scale of this intrinsic variability.  For trampoline, there are significant differences between apparatus, thus we do one regression $\hat{\sigma}_d(c_p)$ per apparatus, shown in Figure~\ref{fig:sigma_APP}. Note that some curves are below 0 for control scores close to 10, but this is extrapolated and there is no scoring data in that range. Marks close to the boundary are very rarely achieved in gymnastics (we have none in our data set) and $\hat{\sigma}_d(c_p)$ is always small compared to the distance to the boundaries. We can therefore disregard the mathematical implications of the bounded marking range.

\begin{table}
\footnotesize
\centering
\begin{tabular}[]{c|cccccc}
	\toprule
	$c_p$ & 7.0 & 7.5 & 8.0 & 8.5 & 9.0 & 9.5 \\ 
	$\hat{\sigma}_d(c_p)$ & 0.31 & 0.27 & 0.24 & 0.20 & 0.15 & 0.09 \\
	\bottomrule
\end{tabular}
\caption{Intrinsic discipline judging error variability: a few numerical values for the fitted standard deviation of judging error as a function of the control score in men's artistic gymnastics.}
\label{tab:magsigma}
\end{table}
This identification of the intrinsic discipline judging error variability allows us to express the judge specific general error $\mu_j$ and the biases ($\beta_{\text{SN}}$, $\beta_{\text{COMP}}$) more precisely. We assume that they have similar properties of heteroscedasticity: intentional and unintentional misjudgements are smaller for the best athletes. Being too far off the panel score is very suspicious and is immediately noticed by officials. We therefore suppose that judges do adapt their scoring behaviour to not stand out too much, no matter if their bias is intentional or not. We can thus assess the individual general tendency $\mu_j$ and the national bias $(\beta_{\text{SN}},\beta_{\text{COMP}})$ as a multiple of the fitted intrinsic discipline judging error standard deviation $\hat{\sigma}_d(c_p)$. Our model becomes  
\begin{equation}\label{eq:standard_formula}
	s_{p,j}=c_p + (\mu_j + \beta_{\text{SN}} \cdot \mathds{1}_{\text{SN}} + \beta_{\text{COMP}} \cdot \mathds{1}_{\text{COMP}})\cdot\hat{\sigma}_d(c_p) +  \epsilon_{p,j}.
      \end{equation}
      
To restrict the number of variables in the regression model, we determine the general judge tendency $\mu_j$ beforehand as
\begin{equation}\label{eq:mu_j}
	\hat{\mu}_j = \frac{1}{n_j} \sum_p \frac{s_{p,j}-c_p}{\hat{\sigma}_d(c_p)}
\end{equation}
which is the average judging error expressed as a multiple of $\hat{\sigma}_d(c_p)$. Since we know $c_p$, $\hat{\mu}_j$ and $\hat{\sigma}_d(c_p)$, we can group them together with the reported judge score $s_{p,j}$ into a single variable. The resulting response variable $d_{p,j}$ is the judging error corrected for the judge specific tendency and given by
\begin{equation}
  \begin{split}
    	\label{eq:1}
	d_{p,j} & \equiv s_{p,j} - c_p - \hat{\mu}_j \cdot \hat{\sigma}_d(c_p) \\ & = ( \beta_{\text{SN}} \cdot \mathds{1}_{\text{SN}} + \beta_{\text{COMP}} \cdot \mathds{1}_{\text{COMP}})\cdot\hat{\sigma}_d(c_p) +  \epsilon_{p,j}.
      \end{split}
    \end{equation}
      
Up to this point we did not specify the distribution of the error term $\epsilon_{p,j}$. We already introduced that the distribution of the judging error for discipline $d$ is heteroscedastic in $c_p$ with mean zero and standard deviation $\hat{\sigma}_d(c_p)$. 
We use further results from our first article \cite{MH2018:gymnastics}
  and calculate a marking score $M_j$ for each judge quantifying his/her particular accuracy as a multiple of $\hat{\sigma}_d(c_p)$. A short introduction to the concept of marking score is presented in Appendix \ref{app:methodology}. We thus model the random error term $\epsilon_{p,j}$ of a specific judge $j$ as a normal random variable with mean zero and standard deviation ${\hat{\sigma}_d(c_p)} \cdot M_j$, i.e., the intrinsic discipline judging error variability multiplied by a judge-specific accuracy factor. The term ${\hat{\sigma}_d(c_p)} \cdot M_j$ is the \emph{intrinsic judging error variability} of a specific judge in discipline $d$. Putting everything together, our final regression model is 
\begin{multline}\label{eq:final_formula}
	d_{p,j}=( \beta_{\text{SN}} \cdot \mathds{1}_{\text{SN}} + \beta_{\text{COMP}} \cdot \mathds{1}_{\text{COMP}})\cdot\hat{\sigma}_d(c_p) +  \epsilon_{p,j} \\
	\text{where} \  \epsilon_{p,j}\sim\mathcal{N}(0,\hat{\sigma}_d^2(c_p) \cdot M_j^2)
\end{multline}
and the parameters can be estimated with the generalized least squares method \cite{Aitken:1936}. We can use Eq.~\eqref{eq:final_formula} to estimate the national bias $(\beta_{\text{COMP}},\beta_{\text{SN}})$ by apparatus, by discipline, by nationality and by judge.

\section{Results and Discussion}
\subsection{National bias by discipline}
Table~\ref{tab:Regression} shows the outcome of the general linear model specified by Eq.~(\ref{eq:final_formula}). The results are split by discipline and stage of the event. While 'All gymnasts' encompasses all performances in the dataset, 'Top 8 finalists' only includes the top eight gymnasts in the final stage of a competition (apparatus and all-around finals). Because the general linear model includes the functional heteroscedasticity variable $\hat{\sigma}_d(c_p)$, the estimated parameters $\beta_{\text{SN}}$ and $\beta_{\text{COMP}}$ are expressed as a multiple of $\hat{\sigma}_d(c_p)$. For instance, $\beta_{\text{SN}}=0.5$ means that the bias level in favor of same-nationality gymnasts is half the intrinsic discipline judging error variability for a specific performance level.
\begin{table}
	\footnotesize
	\centering 
	\def\arraystretch{1.1}
	\setlength{\tabcolsep}{0pt} 
	\begin{tabularx}{\columnwidth}{XlXXXrXrXrlXXrXrXrl} 
		\toprule
		& & & & & \multicolumn{6}{c}{All gymnasts} & & & \multicolumn{6}{c}{Top 8 finalists} \\
		\cline{6-11}\cline{14-19}\\[-1.8ex]
		& & & & & Estimate (se) & & t-stat. & & \multicolumn{2}{c}{p-value} & & & Estimate (se) & & t-stat. & & \multicolumn{2}{c}{p-value} \\ 
		\midrule		
		\multicolumn{19}{l}{\textbf{Acrobatics}}\\
		& $\beta_{\text{SN}}$ & & & & 0.04 (0.06) & & 0.71 & & 0.481 & & & & 0.08 (0.27) & & 0.30 & & 0.767 & \\
		& $\beta_{\text{COMP}}$ & & & & -0.04 (0.03) & & -1.16 & & 0.245 & & & & -0.04 (0.08) & & -0.50 & & 0.615 &\\
		\midrule 
		\multicolumn{19}{l}{\textbf{Aerobics}}\\
		& $\beta_{\text{SN}}$ & & & & 0.25 (0.07) & &  3.65 & & 0.000 &** & & &  0.50 (0.15) & &  3.23 & & 0.001 &** \\
		& $\beta_{\text{COMP}}$ & & & & -0.04 (0.04) & & -1.02 & & 0.307 & & & & -0.01 (0.06) & & -0.16 & & 0.870 &\\
		\midrule 
		\multicolumn{19}{l}{\textbf{Artistics (M)}}\\
		& $\beta_{\text{SN}}$ & & & & 0.43 (0.03) & & 13.31 & & 0.000 &** & & &  0.68 (0.11) & &  6.38 & & 0.000 &** \\
		& $\beta_{\text{COMP}}$ & & & & -0.02 (0.02) & & -1.23 & & 0.217 & & & & -0.05 (0.03) & & -1.61 & & 0.107 &\\
		\midrule
		\multicolumn{19}{l}{\textbf{Artistics (F)}}\\
		& $\beta_{\text{SN}}$ & & & & 0.28 (0.04) & &  6.45 & & 0.000 &** & & & 0.55 (0.13) & & 4.08 & & 0.000 &** \\
		& $\beta_{\text{COMP}}$ & & & & -0.05 (0.02) & & -1.89 & & 0.059 & & & & 0.02 (0.04) & & 0.52 & & 0.601 &\\
		\midrule
		\multicolumn{19}{l}{\textbf{Rhythmics}}\\
		& $\beta_{\text{SN}}$ & & & & 0.34 (0.05) & &  6.96 & & 0.000 &** & & &  0.30 (0.19) & &  1.59 & & 0.112 &\\
		& $\beta_{\text{COMP}}$ & & & & -0.04 (0.03) & & -1.57 & & 0.117 & & & & -0.09 (0.06) & & -1.55 & & 0.122 &\\
		\midrule
		\multicolumn{19}{l}{\textbf{Trampoline}}\\
		& $\beta_{\text{SN}}$ & & & & -0.05 (0.05) & & -0.87 & & 0.387 & & & &  0.02 (0.12) & &  0.13 & & 0.900 &\\
		& $\beta_{\text{COMP}}$ & & & & -0.06 (0.03) & & -1.64 & & 0.101 & & & & -0.11 (0.07) & & -1.68 & & 0.094 &\\
		\midrule
		\multicolumn{19}{l}{\hspace{7mm}Significance code: \ \ $p<0.05$*$,  \ \ p<0.01$**}\\
		\bottomrule 
	\end{tabularx} 
	\caption{Regression results by discipline. Estimated parameters $(\beta_{\text{SN}},\beta_{\text{COMP}})$ indicate the national bias as a multiple of the estimated intrinsic discipline judging error variability $\hat{\sigma}_d(c_p)$. To obtain the nominal bias for a given performance quality $c_p$, multiply the estimated parameter with $\hat{\sigma}_d(c_p)$.}
	\label{tab:Regression} 
\end{table}

The results reveal that in aerobic, artistic and rhythmic gymnastics, judges mark same-nationality gymnasts significantly higher than the other panel judges, whereas national bias is not a systemic issue in acrobatic gymnastics and trampoline. In aerobic and artistic gymnastics in particular, national bias is even more pronounced for finalists than during earlier stages of competitions, in other words judges bend the rules further when it counts. This does not necessarily imply that the nominal bias is higher for the best gymnasts since the intrinsic judging error variability decreases as the performance level improves, but instead that the magnitude of the national bias compared to the other sources of judging errors increases for the best athletes.

The most severe bias appears in men's artistic gymnastics, where judges give gymnasts of the same country an average bonus of almost half the intrinsic discipline judging error variability, and an average bonus of two thirds the intrinsic variability for the best finalists. The best men artistic gymnasts in the world typically get marks between 8.5 and 9.5 depending on the apparatus. Using Figure~\ref{fig:sigma_DIS} or Table~\ref{tab:magsigma}, the national bias $\beta_{\text{SN}}=0.68$ for the top men finalists corresponds to a nominal bias of between 0.06 and 0.15 points depending on the apparatus, or 10\% of the total deductions of the performance. Considering the narrow gaps between the best gymnasts, this is a worrying discovery, both in relative and in absolute terms. We point out again that this is for an average judge; the most biased judges are significantly worse!

The results in Table~\ref{tab:Regression} further show that the penalization of direct competitors of same-nationality athletes is very small in all the disciplines and never reaches a 5\% significance level. This negative bias remains much smaller than the intrinsic judging error variability and therefore has a negligible impact on the final rankings. This is in line with prior research \cite{Ansorge:1988, Popovic:2000}. 

\subsection{National bias by nationality and judge}
\label{sec:bias-by-judge}

\begin{figure}
	\centering
	\includegraphics[width=\columnwidth]{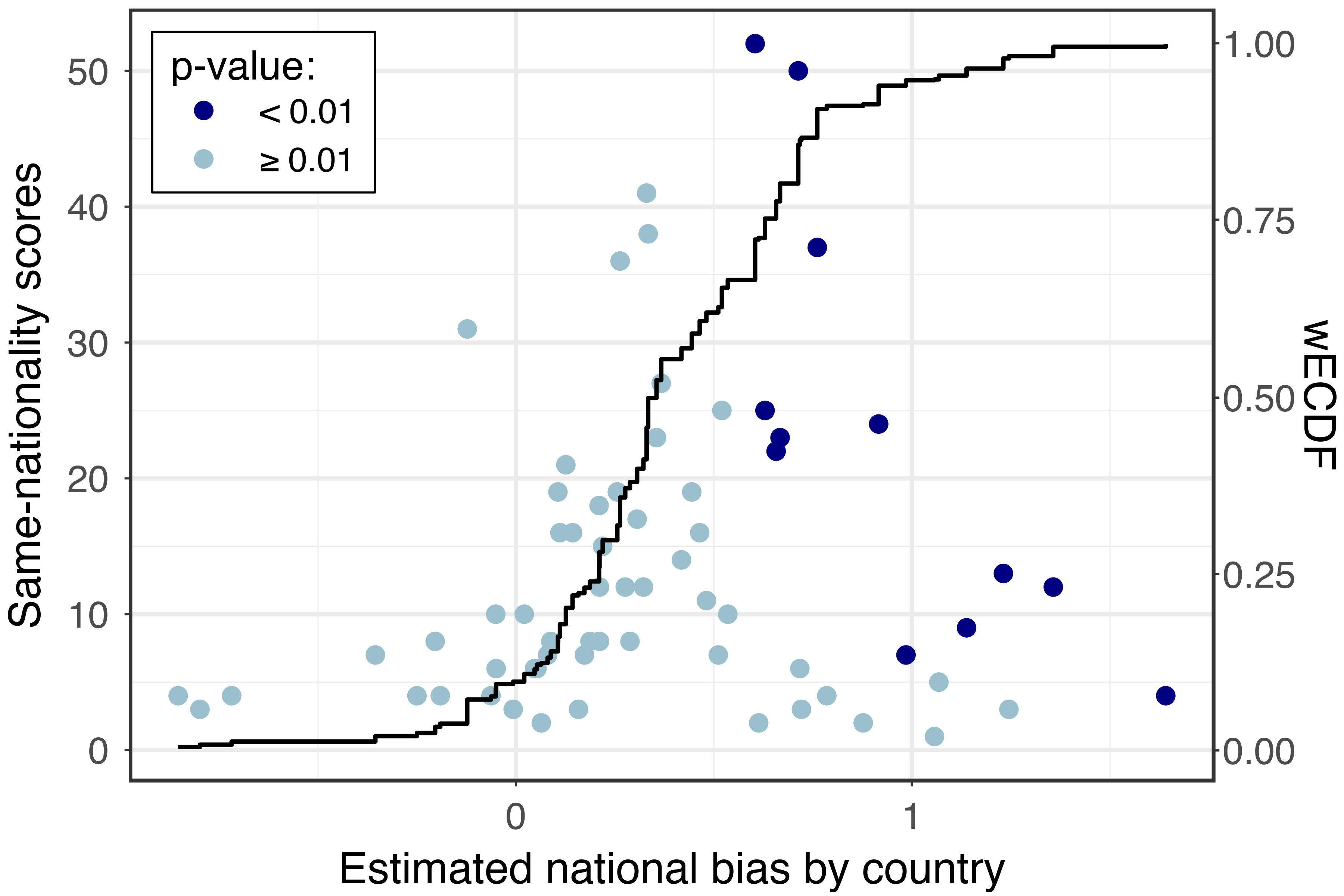}
	\caption{Estimated national bias vs. same-nationality marks by nationality in men's artistic gymnastics together with the weighted empirical cumulative distribution function (wECDF) of the estimations. The statistically significant estimations $(p<0.01)$ are in dark blue.}
	\label{fig:nationality}
\end{figure}
We can apply our general linear regression model by nationality and by judge. We first restrict the number of variables by discarding the penalization of direct competitors, which we have shown to be negligible compared to the positive bias toward same-nationality gymnasts. 

Figure~\ref{fig:nationality} shows the estimated national bias per country against the number of same-nationality marks in men's artistic gymnastics. The national bias is once again expressed as a multiple of the intrinsic discipline judging error variability $\hat{\sigma}_d(c_p)$. The figure also shows the weighted empirical cumulative distribution function (wECDF) by the number of same-nationality marks. We choose this weighting because it best incorporates the higher reliability of the estimation as the number of data points increases. The distribution of the estimated parameter is centered around the magnitude of the national bias $\beta_{\text{SN}}=0.43$. The scatterplot further highlights statistically significant results ($p<0.01$) located in the top right quadrant of the figure. We observe similar results in women's artistic gymnastics and rhythmic gymnastics.

\begin{figure}
	\centering
	\includegraphics[width=\columnwidth]{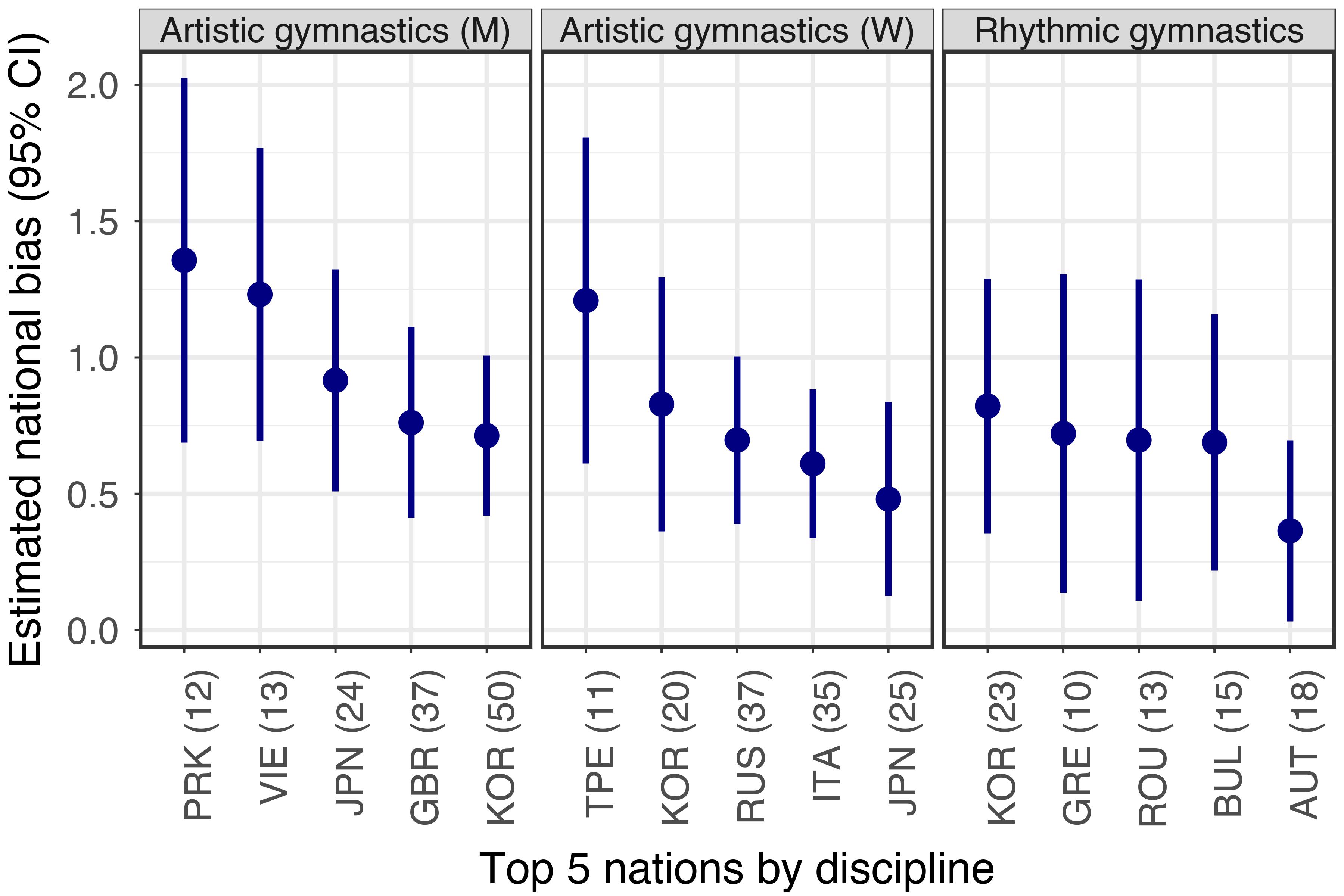}
	\caption{Countries with the highest statistically significant national bias and at least 10 same-nationality marks in artistic gymnastics (men/women) ($p<0.01$)  and rhythmic gymnastics (p<0.05). Number of same-nationality marks in parentheses.}
        \label{fig:Top_Nat}
      \end{figure}

Figure \ref{fig:Top_Nat} shows the nations with the worst statistically significant estimated national bias and at least ten same-nationality marks in men's artistic gymnastics, women's artistic gymnastics, and rhythmic gymnastics. For each included country, the figure shows the 95\% confidence interval of the estimated parameter $\beta_{\text{SN}}$. Eastern European countries dominate the list in rhythmic gymnastics, whereas Asian countries dominate in artistic gymnastics. South Korea is among the worst biased countries in all three disciplines. 
      
\begin{figure}
	\centering
	\includegraphics[width=\columnwidth]{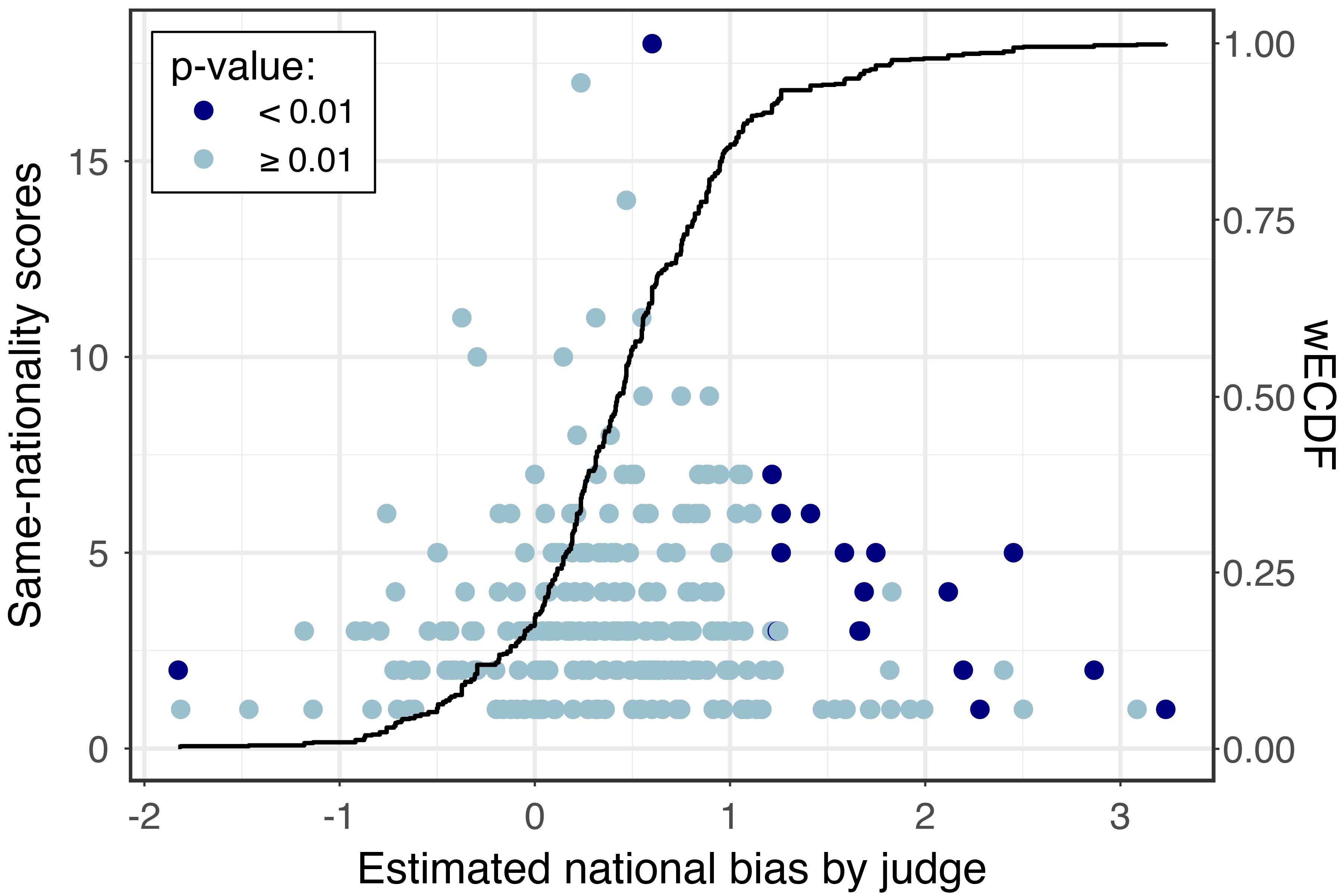}
	\caption{Estimated national bias vs. same-nationality marks by judge in men's artistic gymnastics together with the weighted empirical cumulative distribution function (wECDF) of the estimations. The statistically significant values $(p<0.01)$ are in dark blue.}
	\label{fig:judges}
      \end{figure}

      Figure~\ref{fig:judges} shows the estimated national bias per judge against the number of same-nationality marks in men's artistic gymnastics. The scatterplot also shows the wECDF and the statistically significant results ($p<0.01$). Figure~\ref{fig:judges} exhibits a similar shape as Figure~\ref{fig:nationality}, but with fatter tails and a larger dispersion of the bias. For a judge to exhibit a statically significant national bias, he/she must have shown either a very large bias, or a smaller bias but supported by more data points. There are judges whose national bias is statistically significant and two to three times larger that the intrinsic discipline judging error variability. In other words, their national bias is two and even three times larger than all the sources of errors of an average, fair judge. This is reprehensible, and whether this bias is conscious or not, or caused by other improbable random causes, is irrelevant: these judges should probably be forbidden to ever again judge gymnastics competitions.

\begin{figure}
	\centering
	\includegraphics[width=\columnwidth]{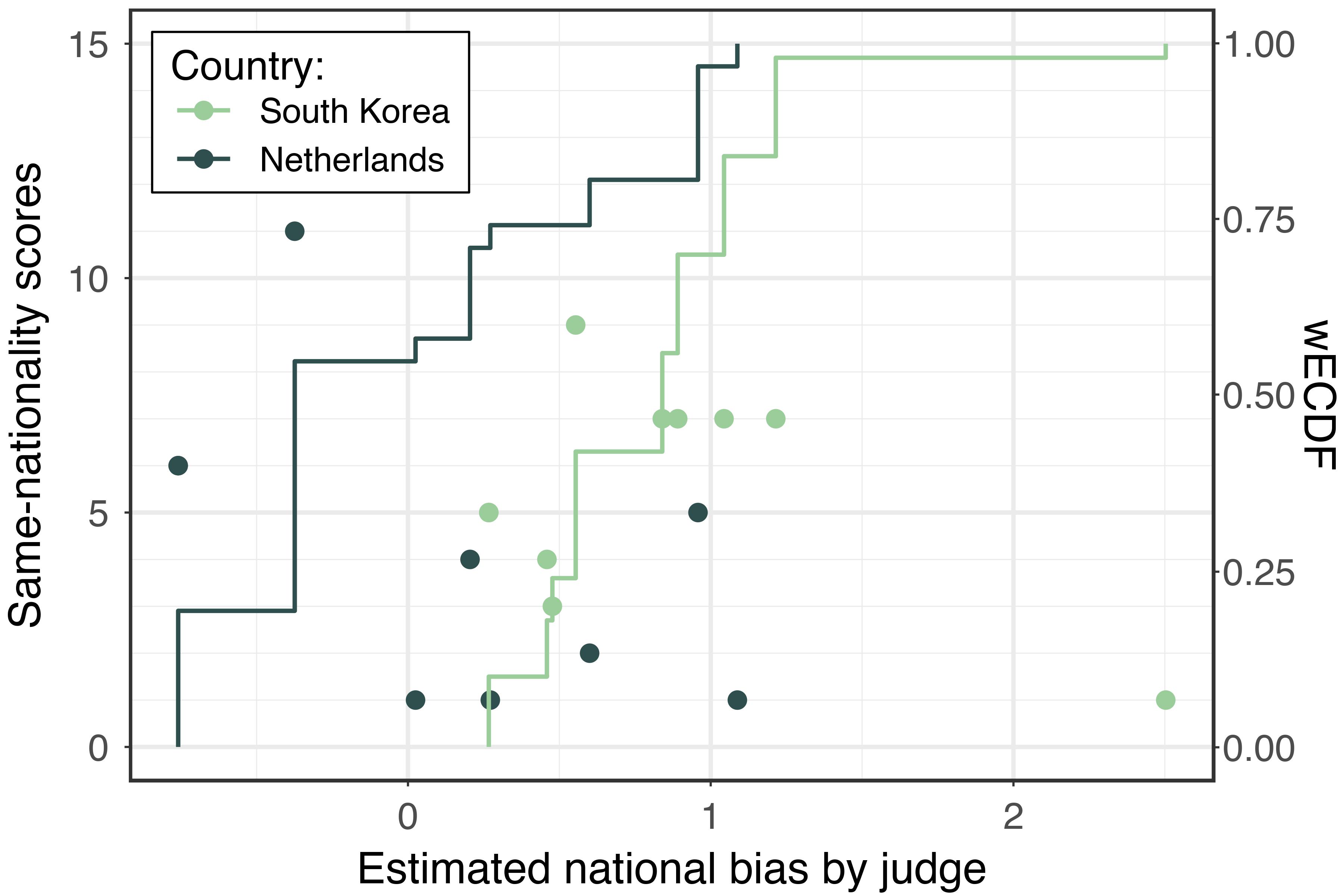}
	\caption{Estimated national bias vs. same-nationality marks of Dutch and South Korean judges in men's artistic gymnastics together with the wECDF of the estimations.}
	\label{fig:ned_kor}
\end{figure}

      We must point out that it is difficult to infer systemic national bias, or lack thereof, based on the average national bias of a country. To illustrate this, Figure~\ref{fig:ned_kor} shows the estimated national bias of all the Dutch and South Korean men's artistic gymnastics judges. Although the two countries have a strikingly different average national bias ($-0.12$ versus $0.71$), both have good and bad judges. The difference can partly be explained by the very large positive bias of a Korean judge and the unusually large negative bias of two Dutch judges.

\subsection{Impact of national bias on rankings}
The regular and significant bias in favor of same-nationality gymnasts naturally raises the question of its impact on the competitions' rankings. We study this further and focus on the apparatus and all-around finals in artistic gymnastics. We apply the scoring aggregation procedure defined in the Codes of Points, and study ranking distortions by calculating the rankings with and without the marks of same-nationality judges. Whenever we observe a same-nationality mark, we discard it and calculate the trimmed mean of the remaining panel marks. If a discarded mark comes from a reference judge, the average reference score simply becomes the mark of the second reference judge. 

In the apparatus finals, 4 out of 740 performances in our dataset include a same-nationality evaluation. For one of the four gymnasts, the final score of the performance is boosted by 0.267 points due to the national bias of a reference judge. The FIG and the other athletes were lucky in this instance because the gymnast in question finished last and far behind the other finalists, and this scoring discordance had no effect on the ranking. In most other instances a difference of this magnitude would have improved the position of the gymnast by a few ranks. This example further illustrates the danger of granting more power to small panels of reference judges. In the all-around finals, evaluations by same-nationality judges are more difficult to avoid and occur in 330 out of 1990 performances in our dataset. In $42\%$ of these all-round finals, removing the potentially biased same-nationality marks changes the final ranking of the gymnasts, including one podium. We propose mitigating measures in the next section.

\subsection{On using the median as the true performance quality}

\label{sec:median}

In our analysis, we use the median panel mark as an approximation of the true performance quality since the FIG did not provide enough post-competition control scores obtained by video review that we could use as anchor points. In the aggregate, this assumption is more than reasonable. The FIG, in numerous discussions, confirmed that most judges are good, and that most judging panels contain a majority of good judges. However, it is possible that a judge flagged as exhibiting a large and statistically significant national bias by our analysis was simply out of consensus with the other panel judges. This is more likely if this judge has only one or two same-nationality evaluations. In these instances, the FIG should carefully review these performances to confirm (or refute) that the median used in the analysis is indeed representative of the true performance quality.

\section{Conclusion and Recommendations}

In this work, we studied the national bias of international gymnastics judges during the 2013--2016 Olympic cycle. The main novelty of our approach over prior work is that we systematically leverage and integrate the intrinsic deviation of judging marks into our national bias regression model. We have shown in \cite{MH2018:gymnastics}
that this intrinsic judging error is heteroscedastic and can be accurately modeled with weighted least-squares exponential regressions. This allows us to express national bias as a function of the intrinsic judging error variability. Intuitively this makes sense: judges are very accurate when evaluating the best gymnasts, thus even a small nominal bias can have a lot of impact on the ranking of the gymnast.

We estimate the national bias by discipline, by nationality and by judge, and show important differences between the least- and most-biased judges. National bias has the largest impact in all-around competitions where it is impossible to avoid same-nationality evaluations.

\subsection{Recommendations}

The FIG has already taken successful steps to decrease the impact of national bias in gymnastics. We now discuss these steps and propose additional mitigation measures.

\subsubsection{Avoid same-nationality judges} The FIG already avoids same-nationality judges in the apparatus finals. This is not possible in all-around finals since they require too many judges.

\subsubsection{Increase the size of the judging panels} 

Same-nationality judging is already difficult to avoid in all-around finals, and larger panels would create even more of these situations. Although large panels are more robust against outliers and could in principle reduce the impact of potentially biased same-nationality judges, we doubt that it is a path worth pursuing. The FIG tries to select the best judges for its most important competitions, and increasing the number of judges in all-around finals might lead to the selection of less accurate judges. Also, the increased number of same-nationality evaluations might encourage suspicious behavior such as vote trading.

\subsubsection{Get rid of the increased power imparted to reference judges} 

In \cite{MH2018:gymnastics}, 
we show that reference judges, who are hand-picked by the FIG and granted more power, are statistically indistinguishable or worse than regular panel judges. We thus recommended that the FIG merges the execution and reference judges into a larger execution panel where all the judges have the same power. The FIG Technical Coordinator is currently proposing the adoption of our recommendation. This will prevent a biased reference judge from potently affecting the final scores of the gymnasts.

\subsubsection{Aggregate marks more aggressively} For all-around finals, where same-nationality evaluations are unavoidable, we recommended that the FIG removes more extreme marks from its panels before aggregating them with a (more aggressive) trimmed mean, and even take the median panel mark. Trampoline already uses the median mark for each jump, but doing the same thing for artistic or rhythmic gymnastics routines would result in more ties. The FIG understandably does not like to award ties, but this should not be an issue in all-around finals since the final ranking includes scores from all the apparatus. Should this unlikely event occur, awarding two gold medals to gymnasts whose performances are within the margin of error of the best judges might not be ideal from a competition perspective, but certainly a better outcome than letting a biased judge act as the tiebreaker. Following our recommendation, the FIG Technical Coordinator is currently proposing to trim the best and worst two marks from its execution panels for all the disciplines except trampoline, which already uses the median. The resulting aggregation would be the average of the middle three marks in artistic gymnastics, and the average of the middle two marks in aerobic, acrobatic and rhythmic gymnastics. This change would apply to all the events, including all-around finals.

\subsubsection{Track the long-term performance of judges and remove the worst culprits} The FIG recently started using an improved Judge Evaluation Program (JEP) to assess the performance of gymnastics judges \cite{MH2018:gymnastics}. 
JEP allows longitudinal monitoring of the judges, many of whom are judging for decades. The bias tool in JEP is not as refined as the analysis done in this article, but it should nevertheless make it easier to identify the most biased judges. 

\subsection{Future work}

Accurate control scores provided by video review post-competitions could allow us to refine our work on national bias. In particular, as discussed in Section~\ref{sec:median}, we could verify that judges exhibiting a large and statistically significant bias were not simply out of consensus with other panel judges. We could also investigate more complex judging behavior in gymnastics such as vote trading and compensation effects revealed in figure skating and ski jumping \cite{Zitzewitz:2006, Zitzewitz:2014}. Further analysis should also include the serial positioning of athletes to better understand the bias against direct competitors of same-nationality gymnasts. Before a same-nationality athlete has performed his/her routine, a judge vaguely knows his/her direct competitors. After the routine, it is more clear who the direct competitors are and to what extent an intentional misjudgement towards them can help the preferred athlete. We expect to see a dependence of the national bias on the positioning of the gymnasts.

National bias has been investigated in many other sports besides gymnastics. In the third article of this series \cite{HM2018:heteroscedasticity},   
we show that the judging error variability in other sports with panels of judges awarding marks within a finite range has a similar heteroscedastic shape. The integration of this behaviour could provide an improved national bias analysis in all these sports.

\section*{Acknowledgments}

We would like to thank Nicolas Buompane, Steve Butcher, André Gueisbuhler, Sylvie Martinet and Rui Vinagre from the FIG, and Christophe Pittet and Pascal Rossier from Longines, for their help, comments and suggestions throughout this work.

\printbibliography

\appendices

\section{Intrinsic discipline judging error variability and marking score}\label{app:methodology}

This appendix summarizes how we estimate the \emph{intrinsic judging error variability} ${\hat{\sigma}_d(c_p)}$ and the \emph{marking score} $M_j$ of judge $j$. More details can be found in \cite{MH2018:gymnastics}.

Let $e_{p,j}=s_{p,j}-c_p$ be the difference between the given mark of judge $j$ for performance~$p$ and the true quality of the performance, indicated by the objective control score $c_p$. We call $e_{p,j}$ the judging error of judge $j$ for performance $p$. We observe that the magnitude of the average judging error has mean 0 and is strongly heteroscedastic: it decreases as the performance level improves. More precisely, the standard deviation of the judging error closely follows an exponential function of the control score $c_p$. We estimate the parameters of the weighted exponential regression model $\sigma_d(c_p)=\alpha_d+\beta_de^{\gamma_dc_p}+\epsilon_d$ based on the data. The weight of each value $c_p$ is its frequency in our data set. This fitted intrinsic judging error variability can be derived by discipline, apparatus, gender, and even by judge.

We then develop a marking score based on the fitted intrinsic judging error variability ${\hat\sigma_d(c_p)}$ to quantify the accuracy of the marks given by the judges. The marking score must be unbiased with respect to the performance quality and the apparatus. This is achieved by normalizing the judging error $e_{p,j}$ by the intrinsic judging error variability $\hat{\sigma}_d(c_p)$ estimated for each discipline/apparatus $d$. The marking score of performance $p$ by judge $j$ is thus given by 
$$m_{p,j}=\frac{e_{p,j}}{\hat{\sigma}_d(c_p)}=\frac{s_{p,j}-c_p}{\hat{\sigma}_d(c_p)}.$$
The overall marking score of a judge $j$ who evaluated $n$ performances is 
$$M_j=\sqrt{\frac{1}{n}\sum_{p=1}^n m_{p,j}^2}.$$ It quantifies the general accuracy of this judge compared to his peers. 
A judge whose every mark deviates from the true performance score by $\hat{\sigma}_d(c_p)$ has a marking score of 1.0, while a perfect judge obtains a marking score of 0. As the intrinsic judging error variability is fitted by discipline, the marking score is comparable between different disciplines. The marking score can also be used to detect outlier marks based on a judge-specific threshold.

\end{document}